\documentclass[twocolumn,showpacs,prl,aps]{revtex4-1}

\usepackage{dcolumn}
\usepackage{amssymb}

\draft

\begin{document}

\title{On origin of 1/f noise in manganites: \\
memoryless transport against %
mysterious slow fluctuators}

\author{Yu. E. Kuzovlev}
\email{kuzovlev@fti.dn.ua} \affiliation{Donetsk
Physics and Technology Institute, 83114 Donetsk, Ukraine}

%\date{\today}

\begin{abstract}
An alternative explanation of 1/f-noise in manganites %
is suggested and discussed.
\end{abstract}


\maketitle

{\bf 1.\,Introduction}.\,\,

The so-called perovskite manganites, or %
colossal magneto-resistance manganites  %
\cite{pm,dag}, are materials known as %
``1/f-noise champions''. For proper references %
see works \cite{d,d0,d1}, since just %
their extremely interesting experimental results %
stimulated my present communication. %
Namely, first, observation of very high level of %
1/f noise in good bulk crystals (instead of thin films %
as usually). %
Second, very weak dependence of this noise %
(expressed in standard relative units %
via\, $\,S_R(f)/R^2\,$) on temperature in wide %
from the room one down to 79$^\circ\,$K\,. %
At 79$^\circ\,$K\, transition to strongly non-ohmic %
regime was found and attracted most %
authors' ``theoretical interest'' %
in \cite{d,d0,d1}.

In my opinion, however, %
discussion of the wide Ohmic region %
may be much more %
useful for understanding nature of 1/f-noise. %
Below I will try to suggest a principal alternative %
to the hypotheses seemingly accepted %
by authors of \cite{d,d0,d1} .

{\bf 2.\,Experimental data}.\,\, %

In the mentioned temperature range, %
resistance of the\, $\,L=0.3\,$mm\, long part of %
rectangular crystal with cross-section %
$\,A=\,$2$\times$3\,mm$^2$\,\, %
changed between\, $\,R(300^o$K$)\approx 2\,$Ohm\, %
and\, $\,R(80^o$K$)\approx 200\,$Ohm\, \cite{d1}. %
At that, the power spectral density (PSD) %
of relative resistance fluctuations was %
practically independent on temperature,
\begin{eqnarray}
\frac {S_V(f)}{V^2}\,=\, \frac {S_R(f)}{R^2}\, %
\approx\, \frac {4\cdot 10^{-11}}f\, \,\, \label{psd}
\end{eqnarray}

{\bf 3.\,Standard interpretation}.\,\, %

Most popular interpretation of %
1/f-noise in solids relates it to some %
hypothetical thermally activated ``fluctuators'' %
with wide enough variety of activation %
energies \cite{dh,w}. %
Under suitable parameters, this model %
can well reproduce both frequency and %
temperature dependencies of 1/-f-type PSDs. %
But it never helped to indicate %
physical nature of ``fluctuators'', %
thus prompting that they hardly exist %
in literal sense.

If, nevertheless, they really take place and, - %
as authors of \cite{d1} do allude, - %
represent more or less local structural %
rearrangements or switchings between %
coexisting phases, then %
we can write
\begin{eqnarray}
\frac {S_V(f)}{V^2}\,=\, \frac {S_R(f)}{R^2}\, %
\sim\, \frac 1{f\,N\,\ln{(f_2/f_1)}}\, \,\,, \label{si}
\end{eqnarray}
where\, $\,f_2\,$ and $\,f_1\,$ are upper and lower %
1/f-noise frequencies under measurements, %
and $\,N\,$ is number of fluctuators %
in the observed volume\, %
$\,\Omega\approx L A\,$ %
(though may be\, $\,\Omega\sim L^3\,$\, %
is more reasonable estimate). %
Comparison of (\ref{si}) with (\ref{psd}) %
gives\, $\,N\sim 10^9\,$\,, and thus %
typical fluctuator takes in a volume %
with linear size\, %
$\,l\sim (\Omega/N)^{1/3} %
\sim 10^{-4}\,$cm\, %
(if not\, $\,l\sim (L^3/N)^{1/3}\sim %
2\cdot 10^{-5}\,$cm\,). %

Likely, that are too large space regions %
to be characterized by so small %
activation energy barriers %
as\, $\,\sim k_B T\,\ln{(f_2/f)}\,$.

{\bf 4.\,Alternative interpretation}.\,\, %

The possibility of coexistence of %
different phases in CMR manganites  %
means that they are materials with %
``strongly correlated electrons'' %
(see e.g. \cite{pm,dag} and %
references in \cite{d,d0,d1}). %
``Strong correlations'' (resulting, in particular, %
from Coulomb interactions and %
Coulomb blockade) %
may strongly decrease effective number %
of free charge carriers, %
i.e. simultaneously and independently %
movable ones %
(see e.g. example in \cite{kmg}). %

The remaining free carriers %
can be considered from viewpoint of another popular %
empirical model \cite{hkv,fnh} where %
\begin{eqnarray}
\frac {S_V(f)}{V^2}\,=\, \frac {S_R(f)}{R^2}\, %
\approx\, \frac {\alpha}{f\,N}\, \,\,, \label{ai}
\end{eqnarray}
with\, $\,N\,$\, being number of carriers %
in the observed volume, and\, $\,\alpha\,$\, %
called ``Hooge constant''. %
In usual crystal materials, %
when inelastic lattice (phonon) %
scattering dominates,\, %
$\,10^{-3}\lesssim\alpha\lesssim 10^{-1}\,$\, %
\cite{hkv,fnh,bk3}. %
%
% with  ``folklore'' value\, %
% $\,\alpha = 2\cdot 10^{-3}\,$. %
%
Taking %
$\,\alpha = 10^{-2}\,$ for rough  %
comparison of (\ref{ai}) and (\ref{psd}), %
we have again\, $\,N\sim 10^9\,$\,, %
but now with movable carriers %
in the role of ``fluctuators''. %

The corresponding %
characteristic length\, %
$\,l\sim (\Omega/N)^{1/3} %
\sim 10^{-4}\,$cm\, seems, of course, %
very large. But, nevertheless, %
it is well compatible with %
the experimental conductivity,\, %
\[
\sigma\,=\, L/RA\,\approx\, %
2.5\cdot 10^{-3}\div 2.5\cdot %
10^{-1}\,\, %
\texttt{Ohm}^{-1}\cdot \texttt{cm}^{-1}\,\,\,, %
\]
if we assume that this conductivity %
is determined by inelastic jumps %
of carriers between relatively isolated %
spatial regions (``grains'') %
with volumes\,$\,\sim\Omega\,$. %

Indeed, if elementary transition through %
a boundary between neighbor regions %
take a time\, $\,\sim\tau\,$\,, %
then maximal (saturation) current per %
one elementary boundary (with area\, $\,\sim l^2\,$) %
is on order of\, $\,J_{max}\sim e/\tau\,$\, %
(the saturation just reflects %
the ``strong correlations''). %
Then in ohmic (low-voltage) regime %
the current must be\, %
\[
J\,\approx\,\frac {eU}{k_B T}\, J_{max}\, %
=\, \frac {e^2U}{k_B T\tau}\,\,\,,
\]
where\, $\,U\approx lV/L\,$\, is potential drop %
across the boundary. %
This means, evidently, %
that ohmic conductivity of such medium %
obeys estimate\, %
\[
\sigma\,\sim\, %
\frac {e^2}{k_B T\tau l}\, \lesssim\, %
\frac {e^2}{\hbar l}\, \sim \,
1\,\,\,\, \texttt{Ohm}^{-1}\cdot \texttt{cm}^{-1}\,\, %
\]
(due to natural restriction\, %
$\,\tau \gtrsim \hbar/k_B T\,$), %
which agrees with above experimental values. %

{\bf 5.\,Free carriers as %
1/f-type fluctuators}.\,\, %

Specific characteristics of the ``grains'' are not %
principally important for low-frequency %
electric noise produced by the free (movable) carriers. %
The only principal thing is that %
the system constantly forgets history of their jumps. %

If it is so, then possible fluctuations in amount %
of charge transport grow with time like %
most probable amount do, %
i.e. nearly proportionally to time %
(for, figuratively speaking, %
the system without memory %
can not distinguish between a  %
``norm'' of transport events and %
their ``excess'' or `1deficiency''). %

This just means that rate of transport %
(PSD of transport noise %
and the system's conductance) undergoes %
scaleless 1/f-type fluctuations %
(so that time-averaged rate varies from one %
experiment to another).
In terms of individual carriers, %
their diffusivities/mobilities %
have no certain value but fluctuate with 1/f-type %
spectrum.

First statistical theory of these fluctuations %
was published in \cite{pjtf,bk12} %
presenting, in particular, clear explanation %
of the ``Hooge constant'' %
(see also \cite{kmg,bk3,i2,p1008,eiphg} %
and references therein). %

{\bf 6.\,Conclusion}.\,\, %

The essence of the appointed view %
is that 1/f-noise comes not from %
hierarchy of long memory times but, %
in opposite, from absence of long %
memory at all. %
Such 1/f-noise is trivially compatible %
with finiteness of ``residence times'' %
of particular carriers in a sample %
(as  well as finiteness of their life-times %
under generation-recombination processes, etc.). %
Thus we eliminate both the corresponding farfetched %
questions \cite{w,fnh} and need in %
mysterious slow ``fluctuators'' inside the sample.

Unfortunately, inertia of scientific prejudices %
is so strong that these simple ideas %
were not assimilateâ during 30 years after %
the works \cite{pjtf,bk12,bk3}. %

The matter is that %
transport processes traditionally are thought  %
as ``stochastic'' ones, in the sense %
of probability theory, with \,{\it a priori}\, %
certain (let numerically unknown) rates. %
But in reality %
they obey the Hamiltonian dynamics  %
which, - as honest %
considerations do show %
\cite{kmg,i2,eiphg,i1,i3,p1,p0802,tmf}, - %
always predicts 1/f fluctuations %
in transport rates. %
Thus, fundamental quantitative theory of 1/f-noise %
in manganites requires, first of all, %
a good Hamiltonian model of charge %
transport in these materials.

%\begin{verbatim}
%\end{verbatim}

%\bibliographystyle{apsrev}
%\bibliography{qftnote}

%\end{document}

%-----------------------------

\end{document}